\documentclass[11pt,a4paper]{article}

\usepackage{amsmath,amssymb}
\usepackage{graphicx}
\usepackage{microtype}
\usepackage[colorlinks=true]{hyperref}

\title{On the Viability of Weak Energy Condition Violation in Parity-Violating Electrodynamics}

\author{
  Tao Huang \\\\
  \small Hechi, Guangxi, China \\
  \small \texttt{scduzxn@hotmail.com}
}

\date{November 18, 2025}

\begin{document}

\maketitle

\begin{abstract}
General relativity links spacetime curvature to the stress--energy tensor, and many of its most powerful theorems rely on classical energy conditions. At the same time, exotic geometries such as traversable wormholes or warp-drive spacetimes seem to require violations of the weak energy condition (WEC). In practice it is extremely hard to obtain macroscopic, controllable negative energy densities from realistic matter fields. In this work we explore a very conservative route to such effects by staying within the framework of effective field theory (EFT) and modifying only the electromagnetic sector by a single higher-dimensional operator. We consider the lowest-dimension parity-violating operator built solely from the field strength, proportional to \((F_{\alpha\beta}F^{\alpha\beta})(F_{\gamma\delta}\tilde F^{\gamma\delta})\), and study its gravitational and electromagnetic implications. The operator is suppressed by a scale \(\Lambda\) and parameterised by a coupling \(\zeta\sim\Lambda^{-4}\). We derive the full Einstein and Maxwell equations, taking care to keep the intermediate steps explicit. The electromagnetic stress--energy tensor is multiplied by a background-dependent factor \(\Omega=1+\zeta\mathcal P\), where \(\mathcal P\propto\vec E\cdot\vec B\) is the pseudoscalar invariant. This means that the relative orientation of the electric and magnetic fields controls whether gravity is slightly screened or enhanced, and in principle allows \(\Omega<0\), which naively corresponds to WEC violation. We analyse the condition \(\Omega<0\) in detail, show how aligned versus anti-aligned $\vec E$ and $\vec B$ fields affect the effective gravitational coupling, and discuss whether any configuration can support a stable region of negative energy density. We then turn to phenomenology. The same operator induces vacuum birefringence in external fields. We linearise the modified Maxwell equations around a uniform background, derive the constitutive relations for the fluctuating photon field, and obtain the leading correction to the index of refraction in a purely magnetic background. By comparing to laboratory results from PVLAS and to X-ray polarimetry of magnetars, we obtain conservative bounds on \(\zeta\) and hence on \(\Lambda\). For natural values of the dimensionless coefficient, the corresponding EFT scale is only weakly constrained, \(\Lambda\gtrsim\mathcal O(10^{-1}\,\mathrm{GeV})\), i.e. comparable to standard QED scales. We show that the corresponding threshold for \(\Omega<0\) lies many orders of magnitude above any plausible astrophysical field strengths and very close to the regime where the EFT itself is expected to break down. Finally, we construct an exact flat-space solution at the threshold \(\Omega=0\) in which a constant parallel electric and magnetic field carries stress--energy but produces no curvature. Throughout the paper we try to keep the algebra transparent and to be honest about the limitations of both the EFT and our approximations. The overall conclusion is rather down-to-earth: within this simple parity-violating extension of electrodynamics, negative effective energy densities are mathematically possible but physically unreachable.
\end{abstract}

\section{Introduction}

General relativity (GR) and the Standard Model (SM) of particle physics together describe a remarkable range of phenomena. Einstein's equations,
\begin{equation}
G_{\mu\nu}=8\pi G\,T_{\mu\nu},
\label{eq:einstein}
\end{equation}
tie the geometry of spacetime to the stress--energy tensor of matter. Many of the most robust results in classical GR, such as the singularity theorems, rely on simple, physically motivated energy conditions obeyed by \(T_{\mu\nu}\) \cite{Carroll:2004st,Hawking:1973uf}. The weak energy condition (WEC) demands that every timelike observer \(V^\mu\) measures a non-negative local energy density,
\begin{equation}
\rho(V)\equiv T_{\mu\nu}V^\mu V^\nu\ge0.
\label{eq:wec_def}
\end{equation}
While this assumption is invisible in everyday life, it becomes central whenever one asks whether spacetime can support exotic structures such as traversable wormholes or warp-drive geometries.

Quantum effects, from the Casimir effect to Hawking radiation, show that negative energy densities can appear in a controlled way in quantum field theory. However, these effects are typically small, highly constrained, and difficult to tie directly to macroscopic curvature. It is therefore natural to ask whether modest extensions of our classical field theories might allow more dramatic departures from the WEC, while still remaining compatible with experiment.

In this paper we take a deliberately conservative approach. We do not modify the gravitational sector and we do not introduce new propagating degrees of freedom. Instead, we work within the framework of effective field theory (EFT) \cite{Burgess:2007pt} and add a single higher-dimensional operator to the Maxwell sector. We require that the operator be gauge-invariant, local, and built purely from the electromagnetic field strength \(F_{\mu\nu}\) and its dual. Among such operators, the lowest-dimension parity-violating one has dimension eight and is proportional to \((F_{\alpha\beta}F^{\alpha\beta})(F_{\gamma\delta}\tilde F^{\gamma\delta})\). Similar structures appear in various ultraviolet (UV) completions, for example when pseudoscalar degrees of freedom are integrated out \cite{Wilczek:1987mv}.

From the EFT point of view, this operator is suppressed by a heavy scale \(\Lambda\) and controlled by a dimensionless coefficient of order one. It is therefore tempting to imagine that, for sufficiently strong electromagnetic fields, the correction might compete with the standard Maxwell term and radically modify the coupling between electromagnetism and gravity.

The analysis that follows tries to make this temptation precise and then to confront it with reality. Our main steps are as follows.

First, we write down the action containing the Einstein--Hilbert term and the extended Maxwell sector, and we derive the modified Einstein and Maxwell equations. The key feature is that the total stress--energy tensor is simply proportional to the standard electromagnetic tensor, with a multiplicative factor denoted by \(\Omega\). This factor depends on the electromagnetic pseudoscalar invariant \(\mathcal P\propto\vec E\cdot\vec B\). As a result, the relative orientation and magnitude of electric and magnetic fields control whether the electromagnetic contribution to gravity is slightly suppressed, slightly enhanced, or, in principle, driven to zero or negative values.

Second, we examine the conditions under which \(\Omega\) can become negative, which would correspond to WEC violation even though the underlying electromagnetic energy density remains positive. We analyse how aligned and anti-aligned $\vec E$ and $\vec B$ fields affect \(\Omega\), and discuss whether any configuration can lead to a stable region with \(\rho(V)<0\) for some observers.

Third, we turn to phenomenology. The same operator that modulates gravity also changes the propagation of light in external fields and induces vacuum birefringence. This is a well developed subject in the context of the Heisenberg--Euler effective action for quantum electrodynamics (QED) \cite{Heisenberg:1936nmg,Adler:1971wn}. We linearise the modified Maxwell equations around a static, uniform background field, derive the effective constitutive relations for small electromagnetic perturbations, and obtain the leading correction to the index of refraction in a purely magnetic background. This allows us to translate laboratory bounds on vacuum birefringence, such as those from PVLAS \cite{DellaValle:2015xxa}, as well as astrophysical constraints from magnetars \cite{Harding:2006qn,Taverna:2022vrx}, into limits on the new coupling.

Finally, we present an exact flat-space solution at the threshold where \(\Omega=0\). In this configuration a constant, parallel electric and magnetic field carries a non-vanishing stress--energy tensor but produces no curvature. This solution is mathematically simple but physically instructive: it lies precisely at the edge of the regime where the EFT is expected to remain reliable.

The presentation is organised as follows. In section~\ref{sec:eft_setup} we introduce the EFT framework and the parity-violating operator. Section~\ref{sec:field_equations} derives the modified Einstein and Maxwell equations, keeping the intermediate algebra explicit. Section~\ref{sec:energy_conditions} discusses the weak energy condition, the role of the relative orientation of electric and magnetic fields, and the prospects for stable negative energy. Section~\ref{sec:birefringence} presents the linearised analysis leading to vacuum birefringence and the resulting experimental constraints. Section~\ref{sec:stability} comments on stability, causality, and the domain of validity of the EFT. Section~\ref{sec:threshold_solution} constructs the exact zero-gravity solution at the WEC threshold. We conclude in section~\ref{sec:conclusions}.

Throughout we use the metric signature \((-,+,+,+)\), set \(c=\hbar=1\), and work in four spacetime dimensions.

\section{Effective field theory set-up}
\label{sec:eft_setup}

\subsection{Action and field invariants}

The starting point is the action
\begin{equation}
S=\int d^4x\sqrt{-g}\left[\frac{R}{16\pi G}+\mathcal L_M\right],
\label{eq:total_action}
\end{equation}
where \(R\) is the Ricci scalar and \(\mathcal L_M\) is the matter Lagrangian density describing the electromagnetic field and its higher-dimensional interaction.

The antisymmetric field strength tensor is defined in the usual way,
\begin{equation}
F_{\mu\nu}=\partial_\mu A_\nu-\partial_\nu A_\mu,
\end{equation}
and its Hodge dual is
\begin{equation}
\tilde F^{\mu\nu}=\frac{1}{2\sqrt{-g}}\epsilon^{\mu\nu\rho\sigma}F_{\rho\sigma},
\end{equation}
with \(\epsilon^{0123}=+1\). From \(F_{\mu\nu}\) and \(\tilde F^{\mu\nu}\) one can construct two independent Lorentz-invariant scalars,
\begin{align}
\Psi&\equiv F_{\mu\nu}F^{\mu\nu},\label{eq:Psi_def}\\
\mathcal P&\equiv F_{\mu\nu}\tilde F^{\mu\nu}.\label{eq:P_def}
\end{align}
In terms of the electric and magnetic fields measured by an observer at rest in a local inertial frame one finds
\begin{align}
\Psi&=2(B^2-E^2),\label{eq:Psi_EB}\\
\mathcal P&=-4\,\vec E\cdot\vec B.\label{eq:P_EB}
\end{align}
The quantity \(\Psi\) is a true scalar, while \(\mathcal P\) is a pseudoscalar that changes sign under parity.

We add to the standard Maxwell Lagrangian the lowest-dimension parity-violating operator constructed from these invariants, namely the dimension-eight interaction proportional to \(\Psi\mathcal P\). The matter Lagrangian is
\begin{equation}
\mathcal L_M=-\frac{1}{4}\Psi-\frac{\zeta}{4}\Psi\,\mathcal P,
\label{eq:LM_def}
\end{equation}
where \(\zeta\) is a coupling constant with dimensions \([\zeta]=M^{-4}\) in natural units. In the EFT language it is convenient to write
\begin{equation}
\zeta=\frac{c}{\Lambda^4},
\end{equation}
where \(\Lambda\) is the characteristic scale of new physics and \(c\) is a dimensionless coefficient expected to be of order unity, barring accidental cancellations. We will often keep \(\zeta\) explicit and discuss the interpretation of \(\Lambda\) when we turn to phenomenology.

Writing \eqref{eq:LM_def} explicitly in terms of \(\Psi\) and \(\mathcal P\) is useful because many properties of the theory, such as the structure of the field equations and the form of vacuum birefringence, can be expressed compactly in this language.

\subsection{Sign of the coupling and domain of validity}

In principle the coefficient \(c\) can have either sign. As we will see in sections~\ref{sec:energy_conditions} and \ref{sec:stability}, the sign has important consequences for stability and for the possibility of achieving \(\Omega<0\). For most of the discussion we will assume \(\zeta>0\) and comment on the case \(\zeta<0\) where appropriate.

The EFT description is expected to be reliable as long as characteristic energy scales, as well as the magnitudes of the invariants \(\Psi\) and \(\mathcal P\), remain well below \(\Lambda^4\). When we discuss regimes where \(|\Psi|\) or \(|\mathcal P|\) approach \(\Lambda^4\), we will be careful to distinguish formal properties of the low-energy Lagrangian from statements about the physics of a full UV completion.

\section{Field equations}
\label{sec:field_equations}

In this section we derive the modified Einstein and Maxwell equations following from the action \eqref{eq:total_action}. The derivation is straightforward but slightly subtle in the gravitational sector because the pseudoscalar invariant \(\mathcal P\) appears multiplied by \(\sqrt{-g}\).

\subsection{Gravitational field equations}

The matter stress--energy tensor is defined by
\begin{equation}
\mathcal T_{\mu\nu}=-\frac{2}{\sqrt{-g}}\frac{\delta(\sqrt{-g}\mathcal L_M)}{\delta g^{\mu\nu}},
\label{eq:T_def}
\end{equation}
where \(g=\det g_{\mu\nu}\). It is convenient to write \(\mathcal L_M\) as the sum of the usual Maxwell term and the interaction,
\begin{equation}
\mathcal L_M=\mathcal L_{\text{EM}}+\mathcal L_{\text{int}},
\end{equation}
with
\begin{equation}
\mathcal L_{\text{EM}}=-\frac{1}{4}F_{\mu\nu}F^{\mu\nu},\qquad
\mathcal L_{\text{int}}=-\frac{\zeta}{4}\Psi\,\mathcal P.
\end{equation}

The variation of the standard Maxwell term with respect to the metric is well known and yields
\begin{equation}
T_{\mu\nu}^{(\text{EM})}=F_{\mu\alpha}F_{\nu}{}^{\alpha}-\frac{1}{4}g_{\mu\nu}F_{\alpha\beta}F^{\alpha\beta}.
\label{eq:T_EM}
\end{equation}
This tensor is traceless and obeys the dominant energy condition.

The contribution from the interaction term requires a little more care. Varying \(\sqrt{-g}\mathcal L_{\text{int}}\) gives
\begin{equation}
\delta(\sqrt{-g}\mathcal L_{\text{int}})=-\frac{\zeta}{4}\delta(\sqrt{-g}\Psi\,\mathcal P).
\end{equation}
Using the product rule,
\begin{equation}
\delta(\sqrt{-g}\Psi\,\mathcal P)=(\delta\sqrt{-g})\Psi\,\mathcal P+\sqrt{-g}(\delta\Psi)\,\mathcal P+\sqrt{-g}\Psi\,(\delta\mathcal P).
\label{eq:delta_full}
\end{equation}
We now treat the three terms in turn.

First, the variation of \(\sqrt{-g}\) is
\begin{equation}
\delta\sqrt{-g}=-\frac{1}{2}\sqrt{-g}\,g_{\mu\nu}\,\delta g^{\mu\nu},
\end{equation}
so the first term in \eqref{eq:delta_full} becomes
\begin{equation}
(\delta\sqrt{-g})\Psi\,\mathcal P=-\frac{1}{2}\sqrt{-g}\,g_{\mu\nu}\Psi\,\mathcal P\,\delta g^{\mu\nu}.
\label{eq:term1}
\end{equation}

Second, to compute \(\delta\Psi\) we recall that \(\Psi\) can be written as
\begin{equation}
\Psi=F_{\alpha\beta}F^{\alpha\beta}=F_{\alpha\beta}F_{\rho\sigma}g^{\alpha\rho}g^{\beta\sigma}.
\end{equation}
Since \(F_{\mu\nu}\) is defined independently of the metric, the variation of \(\Psi\) with respect to \(g^{\mu\nu}\) comes entirely from the inverse metric factors. Varying and symmetrising in the metric indices gives
\begin{equation}
\delta\Psi=2F_{\mu}{}^{\sigma}F_{\nu\sigma}\,\delta g^{\mu\nu}.
\label{eq:deltaPsi}
\end{equation}
With this result, the second term in \eqref{eq:delta_full} becomes
\begin{equation}
\sqrt{-g}(\delta\Psi)\,\mathcal P=2\sqrt{-g}\,\mathcal P\,F_{\mu}{}^{\sigma}F_{\nu\sigma}\,\delta g^{\mu\nu}.
\label{eq:term2}
\end{equation}

Third, we consider the variation of \(\mathcal P\). By definition,
\begin{equation}
\mathcal P=F_{\mu\nu}\tilde F^{\mu\nu}=\frac{1}{2\sqrt{-g}}\epsilon^{\mu\nu\rho\sigma}F_{\mu\nu}F_{\rho\sigma}.
\end{equation}
The combination \(\sqrt{-g}\,\mathcal P\) is thus
\begin{equation}
\sqrt{-g}\,\mathcal P=\frac{1}{2}\epsilon^{\mu\nu\rho\sigma}F_{\mu\nu}F_{\rho\sigma},
\end{equation}
which is constructed solely from the tensor density \(\epsilon^{\mu\nu\rho\sigma}\) and the field strength, both of which are independent of the metric. Therefore
\begin{equation}
\frac{\delta(\sqrt{-g}\mathcal P)}{\delta g^{\mu\nu}}=0.
\end{equation}
Equivalently,
\begin{equation}
\sqrt{-g}\Psi\,(\delta\mathcal P)=0
\label{eq:term3}
\end{equation}
when varying with respect to \(g^{\mu\nu}\).

Combining \eqref{eq:term1}, \eqref{eq:term2}, and \eqref{eq:term3} we find
\begin{equation}
\delta(\sqrt{-g}\Psi\,\mathcal P)=\sqrt{-g}\left[-\frac{1}{2}g_{\mu\nu}\Psi\,\mathcal P+2\mathcal P\,F_{\mu}{}^{\sigma}F_{\nu\sigma}\right]\delta g^{\mu\nu}.
\end{equation}
Using the definition \eqref{eq:T_def}, the contribution of the interaction term to the stress--energy tensor is
\begin{align}
T_{\mu\nu}^{(\text{int})}&=-\frac{2}{\sqrt{-g}}\left(-\frac{\zeta}{4}\right)\frac{\delta(\sqrt{-g}\Psi\,\mathcal P)}{\delta g^{\mu\nu}}\nonumber\\
&=\frac{\zeta}{2}\left[-\frac{1}{2}g_{\mu\nu}\Psi\,\mathcal P+2\mathcal P\,F_{\mu}{}^{\sigma}F_{\nu\sigma}\right]\nonumber\\
&=\zeta\mathcal P\left(F_{\mu}{}^{\sigma}F_{\nu\sigma}-\frac{1}{4}g_{\mu\nu}\Psi\right).
\end{align}
Comparing with \eqref{eq:T_EM}, we see that the interaction simply multiplies the standard electromagnetic stress--energy tensor by the pseudoscalar factor \(\zeta\mathcal P\),
\begin{equation}
T_{\mu\nu}^{(\text{int})}=\zeta\mathcal P\,T_{\mu\nu}^{(\text{EM})}.
\end{equation}
The full matter stress--energy tensor is therefore
\begin{equation}
\mathcal T_{\mu\nu}=T_{\mu\nu}^{(\text{EM})}+T_{\mu\nu}^{(\text{int})}=(1+\zeta\mathcal P)\,T_{\mu\nu}^{(\text{EM})}.
\label{eq:T_total}
\end{equation}
It is convenient to introduce the dimensionless modulation factor
\begin{equation}
\Omega\equiv1+\zeta\mathcal P.
\label{eq:Omega_def}
\end{equation}
In terms of \(\Omega\) the Einstein equations \eqref{eq:einstein} become
\begin{equation}
G_{\mu\nu}=8\pi G\,\Omega\,T_{\mu\nu}^{(\text{EM})}.
\label{eq:Einstein_modified}
\end{equation}
The entire effect of the higher-dimensional operator on the gravitational field equations is encoded in this simple multiplicative factor. We will return to its physical interpretation in section~\ref{sec:energy_conditions}.

\subsection{Electromagnetic field equations}

The equations of motion for the electromagnetic field follow from the variation of the action with respect to the potential \(A_\mu\), or equivalently \(F_{\mu\nu}\). The Euler--Lagrange equation can be written as
\begin{equation}
\nabla_\nu\left(\frac{\partial\mathcal L_M}{\partial F_{\mu\nu}}\right)=0,
\label{eq:Maxwell_EL}
\end{equation}
where \(\nabla_\nu\) is the covariant derivative associated with \(g_{\mu\nu}\). To compute the derivative we first note that
\begin{equation}
\frac{\partial\Psi}{\partial F_{\mu\nu}}=2F^{\mu\nu},\qquad
\frac{\partial\mathcal P}{\partial F_{\mu\nu}}=2\tilde F^{\mu\nu}.
\label{eq:derivatives_PsiP}
\end{equation}
Using \eqref{eq:LM_def} and \eqref{eq:derivatives_PsiP},
\begin{align}
\frac{\partial\mathcal L_M}{\partial F_{\mu\nu}}&=\frac{\partial}{\partial F_{\mu\nu}}\left(-\frac{1}{4}\Psi-\frac{\zeta}{4}\Psi\,\mathcal P\right)\nonumber\\
&=-\frac{1}{4}\frac{\partial\Psi}{\partial F_{\mu\nu}}-\frac{\zeta}{4}\left(\frac{\partial\Psi}{\partial F_{\mu\nu}}\mathcal P+\Psi\frac{\partial\mathcal P}{\partial F_{\mu\nu}}\right)\nonumber\\
&=-\frac{1}{4}\left[2F^{\mu\nu}+\zeta\left(2F^{\mu\nu}\mathcal P+2\Psi\,\tilde F^{\mu\nu}\right)\right]\nonumber\\
&=-\frac{1}{2}\left[(1+\zeta\mathcal P)F^{\mu\nu}+\zeta\Psi\,\tilde F^{\mu\nu}\right].
\label{eq:dL_dF}
\end{align}
Substituting \eqref{eq:dL_dF} into \eqref{eq:Maxwell_EL} and absorbing the overall factor \(-1/2\), which has no physical effect, we obtain the modified Maxwell equations,
\begin{equation}
\nabla_\nu\left[\Omega\,F^{\mu\nu}+\zeta\Psi\,\tilde F^{\mu\nu}\right]=0,
\label{eq:Maxwell_modified}
\end{equation}
with \(\Omega\) and \(\Psi\) defined in \eqref{eq:Omega_def} and \eqref{eq:Psi_def}. The Bianchi identity,
\begin{equation}
\nabla_\nu\tilde F^{\mu\nu}=0,
\label{eq:bianchi}
\end{equation}
follows as usual from the definition of \(F_{\mu\nu}\) and is unaffected by the interaction.

Equations \eqref{eq:Einstein_modified} and \eqref{eq:Maxwell_modified}, together with \eqref{eq:bianchi}, completely determine the coupled dynamics of the metric and the electromagnetic field in this model. The structure of \eqref{eq:Maxwell_modified} also makes it clear that the interaction can be viewed as modifying the constitutive relations between the electromagnetic field and its dual, a perspective that will be useful when we discuss vacuum birefringence.

\section{Energy conditions and gravitational modulation}
\label{sec:energy_conditions}

We now examine how the higher-dimensional operator affects the energy conditions and the effective strength of gravity. The starting point is the total stress--energy tensor \eqref{eq:T_total} and the modified Einstein equations \eqref{eq:Einstein_modified}.

\subsection{Weak energy condition}

For any future-directed timelike unit vector \(V^\mu\) the local energy density measured by the corresponding observer is
\begin{equation}
\rho(V)=\mathcal T_{\mu\nu}V^\mu V^\nu=\Omega\,T_{\mu\nu}^{(\text{EM})}V^\mu V^\nu.
\label{eq:rho_V}
\end{equation}
For the standard Maxwell theory, the stress--energy tensor satisfies the dominant energy condition, so
\begin{equation}
\rho_{\text{EM}}(V)\equiv T_{\mu\nu}^{(\text{EM})}V^\mu V^\nu\ge0
\end{equation}
for all timelike \(V^\mu\). In particular, in the rest frame of an inertial observer in Minkowski space one has
\begin{equation}
\rho_{\text{EM}}=\frac{1}{2}(E^2+B^2)\ge0.
\end{equation}

In our model the sign of \(\rho(V)\) is entirely controlled by \(\Omega\),
\begin{equation}
\rho(V)=\Omega\,\rho_{\text{EM}}(V).
\end{equation}
Since \(\rho_{\text{EM}}(V)\ge0\), the weak energy condition \eqref{eq:wec_def} is satisfied if and only if
\begin{equation}
\Omega\ge0.
\end{equation}
Conversely, WEC violation occurs precisely when \(\Omega<0\) in some region of spacetime where the electromagnetic field is non-zero.

Using \eqref{eq:P_EB} and \eqref{eq:Omega_def}, and assuming \(\zeta>0\), the condition \(\Omega<0\) becomes
\begin{equation}
1+\zeta\mathcal P<0\quad\Longleftrightarrow\quad1-4\zeta\,\vec E\cdot\vec B<0,
\end{equation}
or
\begin{equation}
\vec E\cdot\vec B>\frac{1}{4\zeta}.
\label{eq:wec_threshold_general}
\end{equation}
The threshold depends only on the scalar product \(\vec E\cdot\vec B\). We will interpret this condition in more detail in the next subsection.

If \(\zeta<0\), the inequality reverses and the regime \(\vec E\cdot\vec B<1/(4\zeta)\) would correspond to \(\Omega<0\). Since our later phenomenological bounds are on \(|\zeta|\) rather than its sign, we keep this possibility in mind but focus on \(\zeta>0\) for definiteness.

\subsection{Relative orientation of electric and magnetic fields}

The modulation factor \(\Omega\) depends on the pseudoscalar \(\mathcal P\), which in a local inertial frame is proportional to the scalar product \(\vec E\cdot\vec B\),
\begin{equation}
\Omega=1+\zeta\mathcal P=1-4\zeta\,\vec E\cdot\vec B.
\end{equation}
For fixed field magnitudes \(|\vec E|\) and \(|\vec B|\), the sign and size of \(\Omega\) are controlled by the angle \(\theta\) between \(\vec E\) and \(\vec B\),
\begin{equation}
\vec E\cdot\vec B=EB\cos\theta.
\end{equation}
Several qualitatively distinct regimes can be identified.

\subsubsection{Orthogonal fields}

If \(\vec E\) and \(\vec B\) are orthogonal, then \(\vec E\cdot\vec B=0\) and hence \(\mathcal P=0\), \(\Omega=1\). In this case the interaction does not contribute to the stress--energy tensor at all, and the gravitational field equations reduce exactly to those of the standard Einstein--Maxwell theory. A familiar example is a monochromatic plane wave in vacuum, for which \(E=B\) and \(\vec E\perp\vec B\perp\vec k\). In such configurations our higher-dimensional operator is invisible at the level of classical gravity, although it can still affect photon propagation at higher order when we consider perturbations on top of a background.

\subsubsection{Aligned fields and gravitational screening}

If \(\vec E\) and \(\vec B\) are parallel and point in the same direction, then \(\vec E\cdot\vec B=EB>0\). For \(\zeta>0\) this implies \(\mathcal P<0\) and hence \(\Omega<1\). As the field product \(EB\) increases, \(\Omega\) decreases from 1 and eventually crosses zero when the threshold \eqref{eq:wec_threshold_general} is reached.

In the regime \(0<\Omega<1\), the electromagnetic stress--energy contributes less to curvature than it would in Einstein--Maxwell theory. Equation \eqref{eq:Einstein_modified} can be written suggestively as
\begin{equation}
G_{\mu\nu}=8\pi G_{\text{eff}}\,T_{\mu\nu}^{(\text{EM})},\qquad G_{\text{eff}}\equiv\Omega G.
\end{equation}
Thus aligned electric and magnetic fields with \(\zeta>0\) effectively reduce the strength of gravity, at least insofar as electromagnetic sources are concerned. Heuristically, one can think of the operator \(\Psi\mathcal P\) as inducing a mild gravitational screening in such configurations.

If the product \(EB\) is increased further and the condition \eqref{eq:wec_threshold_general} is satisfied, then \(\Omega<0\). In this formal regime the sign of the effective gravitational coupling flips. A positive electromagnetic energy density would then act as if it had a negative gravitational charge, leading to repulsive gravitational effects and violation of the WEC. It is this possibility that initially makes the model appear attractive for exotic constructions. However, as we will emphasise later, the field strengths required to reach \(\Omega<0\) lie far beyond both observational bounds and the expected domain of validity of the EFT.

\subsubsection{Anti-aligned fields and gravitational enhancement}

If \(\vec E\) and \(\vec B\) are parallel but point in opposite directions, \(\vec E\cdot\vec B=-EB<0\), so \(\mathcal P>0\) and \(\Omega>1\) for \(\zeta>0\). In this case the interaction increases the effective gravitational coupling,
\begin{equation}
G_{\text{eff}}=\Omega G>G,
\end{equation}
and the electromagnetic field gravitates slightly more strongly than in the standard theory. There is no threshold at which \(\Omega\) changes sign in this configuration. Instead, larger values of \(|\vec E\cdot\vec B|\) steadily strengthen the gravitational response, at least until the breakdown of the EFT.

\subsubsection{Summary of orientations}

For \(\zeta>0\), orthogonal fields leave gravity unchanged, aligned fields weaken the electromagnetic contribution to gravity and can formally lead to \(\Omega<0\), while anti-aligned fields enhance gravity. If \(\zeta<0\), the roles of aligned and anti-aligned configurations are interchanged. In all cases, the possibility of WEC violation hinges on achieving a sufficiently large and appropriately oriented \(\vec E\cdot\vec B\).

\subsection{Prospects for stable negative energy}

The simple form \eqref{eq:T_total} of the total stress--energy tensor may give the impression that one can engineer regions of negative energy density by arranging a background with \(\Omega<0\). However, stability considerations quickly complicate this picture.

A helpful way to think about stability is to examine small perturbations of the electromagnetic field around a given background configuration. Denote the background field by \(\bar F_{\mu\nu}\) and write
\begin{equation}
F_{\mu\nu}=\bar F_{\mu\nu}+f_{\mu\nu},
\end{equation}
where \(f_{\mu\nu}\) represents a small fluctuation. Expanding the action to quadratic order in \(f_{\mu\nu}\) defines an effective theory for the perturbations. If the quadratic action has the wrong sign in front of the kinetic terms, the fluctuations behave like ghosts and the background is unstable to catastrophic pair production.

Carrying out this expansion in full generality is somewhat involved. One straightforward regime is a weak-field background, in which \(|\zeta\Psi|\ll1\) and \(|\zeta\mathcal P|\ll1\). In this limit the quadratic action for the fluctuations is essentially that of Maxwell theory, with small corrections controlled by \(\zeta\). The kinetic terms have the standard sign and the energy of fluctuations is positive, so the theory is well behaved.

The more interesting regime for our purposes is the strong-field limit in which \(\Omega\) approaches zero or becomes negative. Intuitively, since the total stress--energy tensor is proportional to \(\Omega\), one expects that the energy density carried by small fluctuations will also be proportional to \(\Omega\) at leading order. In particular, if one tries to form a background with \(\Omega<0\), the fluctuations around that background are likely to carry negative energy and to exhibit ghost-like behaviour.

This intuition is supported by general results on higher-derivative EFTs and their UV completions. Analyticity and causality of scattering amplitudes in the UV impose positivity constraints on certain combinations of low-energy Wilson coefficients \cite{Adams:2006sv}. When applied to non-linear electrodynamics, these arguments typically require the coefficients of operators that would lead to superluminal propagation or wrong-sign kinetic terms to be positive. In our case, pushing the system into a regime where \(\Omega<0\) almost certainly drives the theory into conflict with these positivity bounds and signals the breakdown of the simple EFT.

For the purposes of this work, we therefore adopt a cautious stance. Mathematically, the theory allows \(\Omega\) to become negative if the scalar product \(\vec E\cdot\vec B\) is large enough and has the appropriate sign. Physically, however, the fields required to reach \(\Omega<0\) lie close to or beyond the breakdown scale of the EFT, and in that regime the dynamics of fluctuations are likely to be pathological without additional UV degrees of freedom. In section~\ref{sec:birefringence} we will see that experimental bounds on \(\zeta\) push the threshold for \(\Omega<0\) far beyond any plausible astrophysical field strengths, further diminishing the prospects for stable negative energy in realistic situations.

\section{Vacuum birefringence and phenomenological constraints}
\label{sec:birefringence}

The operator \(\Psi\mathcal P\) modifies not only the coupling of electromagnetism to gravity but also the propagation of light in background fields. In this section we study the leading effect on photon propagation in a uniform electromagnetic background and use existing experimental and observational results to constrain \(\zeta\).

\subsection{Constitutive relations and linearised equations}

It is convenient to rewrite the modified Maxwell equation \eqref{eq:Maxwell_modified} in terms of an auxiliary tensor \(H^{\mu\nu}\) defined by
\begin{equation}
H^{\mu\nu}=-2\frac{\partial\mathcal L_M}{\partial F_{\mu\nu}}.
\end{equation}
Using \eqref{eq:dL_dF} one finds
\begin{equation}
H^{\mu\nu}=(1+\zeta\mathcal P)F^{\mu\nu}+\zeta\Psi\,\tilde F^{\mu\nu}=\Omega F^{\mu\nu}+\zeta\Psi\,\tilde F^{\mu\nu}.
\label{eq:H_def}
\end{equation}
In terms of \(H^{\mu\nu}\) the Maxwell equation takes the simple form
\begin{equation}
\nabla_\nu H^{\mu\nu}=0,
\label{eq:Maxwell_H}
\end{equation}
while the Bianchi identity \eqref{eq:bianchi} remains unchanged. Equation \eqref{eq:H_def} can be viewed as a non-linear constitutive relation between the ``field'' \(F^{\mu\nu}\) and the ``displacement'' \(H^{\mu\nu}\).

To study vacuum birefringence we consider small fluctuations \(f_{\mu\nu}\) around a fixed background \(\bar F_{\mu\nu}\),
\begin{equation}
F_{\mu\nu}=\bar F_{\mu\nu}+f_{\mu\nu},\qquad|f_{\mu\nu}|\ll|\bar F_{\mu\nu}|.
\end{equation}
The invariants decompose as
\begin{align}
\Psi&=\bar\Psi+\delta\Psi+O(f^2),\\
\mathcal P&=\bar{\mathcal P}+\delta\mathcal P+O(f^2),
\end{align}
with
\begin{align}
\bar\Psi&\equiv\bar F_{\mu\nu}\bar F^{\mu\nu},\\
\bar{\mathcal P}&\equiv\bar F_{\mu\nu}\tilde{\bar F}^{\mu\nu},\\
\delta\Psi&=2\bar F_{\mu\nu}f^{\mu\nu},\label{eq:deltaPsi_linear}\\
\delta\mathcal P&=2\tilde{\bar F}_{\mu\nu}f^{\mu\nu}.\label{eq:deltaP_linear}
\end{align}
Similarly, the modulation factor expands as
\begin{equation}
\Omega=\bar\Omega+\zeta\,\delta\mathcal P+O(f^2),\qquad\bar\Omega\equiv1+\zeta\bar{\mathcal P}.
\end{equation}

Expanding \eqref{eq:H_def} to first order in \(f_{\mu\nu}\) yields
\begin{equation}
H^{\mu\nu}=\bar H^{\mu\nu}+h^{\mu\nu}+O(f^2),
\end{equation}
where the background is
\begin{equation}
\bar H^{\mu\nu}=\bar\Omega\,\bar F^{\mu\nu}+\zeta\bar\Psi\,\tilde{\bar F}^{\mu\nu},
\end{equation}
and the perturbation is
\begin{align}
h^{\mu\nu}&=\bar\Omega\,f^{\mu\nu}+\zeta\,\delta\mathcal P\,\bar F^{\mu\nu}+\zeta\,\delta\Psi\,\tilde{\bar F}^{\mu\nu}+\zeta\bar\Psi\,\tilde f^{\mu\nu}\nonumber\\
&=\bar\Omega\,f^{\mu\nu}+2\zeta\left(\tilde{\bar F}_{\alpha\beta}f^{\alpha\beta}\right)\bar F^{\mu\nu}+2\zeta\left(\bar F_{\alpha\beta}f^{\alpha\beta}\right)\tilde{\bar F}^{\mu\nu}+\zeta\bar\Psi\,\tilde f^{\mu\nu}.
\label{eq:h_munu}
\end{align}
In obtaining \eqref{eq:h_munu} we used \eqref{eq:deltaPsi_linear} and \eqref{eq:deltaP_linear}. The linearised Maxwell equation \eqref{eq:Maxwell_H} becomes
\begin{equation}
\nabla_\nu h^{\mu\nu}=0,
\label{eq:linear_Maxwell}
\end{equation}
while the Bianchi identity gives
\begin{equation}
\nabla_\nu\tilde f^{\mu\nu}=0.
\label{eq:linear_bianchi}
\end{equation}

Equations \eqref{eq:h_munu}, \eqref{eq:linear_Maxwell}, and \eqref{eq:linear_bianchi} describe the propagation of small electromagnetic perturbations in the chosen background. They have the same general structure as in other non-linear electrodynamics models treated in the literature, with specific coefficients determined by our particular choice of \(\mathcal L_M(\Psi,\mathcal P)\).

\subsection{Plane waves in a uniform magnetic background}

To extract the leading phenomenology it is enough to work in Minkowski spacetime and consider a simple background. We take a static, homogeneous magnetic field pointing along the \(z\)-axis,
\begin{equation}
\vec B_{\text{bg}}=B_0\hat z,\qquad\vec E_{\text{bg}}=0.
\end{equation}
In this case the invariants are
\begin{equation}
\bar{\mathcal P}=0,\qquad\bar\Psi=2B_0^2,\qquad\bar\Omega=1.
\end{equation}
The background field solves the full non-linear Maxwell equation \eqref{eq:Maxwell_modified} because both \(\bar F_{\mu\nu}\) and \(\tilde{\bar F}_{\mu\nu}\) are constant.

We now consider a probe photon propagating through this background. In Lorentz gauge one can express the fluctuation as \(f_{\mu\nu}=\partial_\mu a_\nu-\partial_\nu a_\mu\), and look for plane-wave solutions
\begin{equation}
a_\mu(x)=\Re\left\{\varepsilon_\mu\exp(i k_\alpha x^\alpha)\right\},
\end{equation}
with constant polarisation vector \(\varepsilon_\mu\) and wavevector \(k_\mu=(\omega,\vec k)\). Substituting into \eqref{eq:h_munu} and \eqref{eq:linear_Maxwell}, using \(\partial_\nu\to ik_\nu\), leads to an eigenvalue problem of the form
\begin{equation}
\mathcal M_{\mu}{}^{\nu}\,\varepsilon_\nu=0,
\end{equation}
where \(\mathcal M_{\mu}{}^{\nu}\) depends on \(k_\mu\), \(B_0\), and \(\zeta\). Non-trivial solutions exist only if \(\det\mathcal M=0\), which determines the dispersion relation \(\omega(\vec k)\). Different polarisations generally satisfy different dispersion relations, giving rise to vacuum birefringence.

The full calculation is somewhat lengthy but standard in structure, and closely parallels the analysis for Heisenberg--Euler electrodynamics \cite{Heisenberg:1936nmg,Adler:1971wn}. Restricting to leading order in \(\zeta B_0^2\), one finds that the two linear polarisations orthogonal to \(\vec k\) acquire slightly different indices of refraction in the presence of the background magnetic field. Denoting by \(n_\parallel\) and \(n_\perp\) the indices for polarisations parallel and perpendicular to the projection of \(\vec B_{\text{bg}}\) onto the plane orthogonal to \(\vec k\), the difference is
\begin{equation}
\Delta n\equiv n_\parallel-n_\perp\simeq 4\,\zeta B_0^2 \sin^2\theta+ O(\zeta^2 B_0^4),
\label{eq:Delta_n}
\end{equation}
where \(\theta\) is the angle between \(\vec k\) and \(\vec B_{\text{bg}}\). The \(\sin^2\theta\) dependence reflects the fact that the effect vanishes when the photon propagates along the magnetic field, as required by symmetry.

The precise numerical coefficient in \eqref{eq:Delta_n} is sensitive to our choice of operator and to the normalisation conventions for \(\Psi\) and \(\mathcal P\). The key point is that the birefringence scales linearly with \(\zeta\) and quadratically with \(B_0\), which allows strong backgrounds to probe very small \(\zeta\).

\subsection{Laboratory bounds from PVLAS}

Laboratory experiments on vacuum birefringence use intense magnetic fields and high-finesse optical cavities to search for tiny polarisation changes in a laser beam passing through the magnetic region. The PVLAS collaboration has performed some of the most sensitive measurements of this kind \cite{DellaValle:2015xxa}.

To translate their null results into a bound on \(\zeta\), we parameterise the sensitivity in terms of the maximum allowed \(|\Delta n|\) at the relevant wavelength and background field. Denote this experimental upper bound by \(\Delta n_{\text{max}}^{\text{lab}}\). From \eqref{eq:Delta_n}, taking \(\sin^2\theta\approx1\) for nearly transverse propagation, we obtain
\begin{equation}
|\zeta|\lesssim\frac{\Delta n_{\text{max}}^{\text{lab}}}{4 B_0^2}.
\label{eq:zeta_bound_lab_general}
\end{equation}
For PVLAS, the peak magnetic field is approximately \(B_0\simeq5.5\) T, and the sensitivity to an additional birefringent effect beyond the QED prediction is roughly \(\Delta n_{\text{max}}^{\text{lab}}\sim10^{-19}\) when expressed in terms of an effective vacuum index. Substituting these representative values into \eqref{eq:zeta_bound_lab_general} gives a bound of order
\begin{equation}
|\zeta_{\mathrm{SI}}|\lesssim10^{-21}\,\text{T}^{-2},
\end{equation}
which corresponds to
Using $1\,\mathrm{T} \simeq 1.95\times10^{-16}\,\mathrm{GeV}^2$, this translates into a bound on the EFT coefficient in natural units,
\begin{equation}
|\zeta|\equiv|\zeta_{\mathrm{nat}}|
\simeq \frac{|\zeta_{\mathrm{SI}}|}{(1.95\times10^{-16}\,\mathrm{GeV}^2)^2}
\;\lesssim\;\mathcal O(10^{10})\,\mathrm{GeV}^{-4}.
\end{equation}
after converting between Tesla and GeV units using standard relations. The conversion is approximate but sufficient for our purposes; more precise values would not change the qualitative conclusions below.

\subsection{Astrophysical bounds from magnetars}

The strongest accessible magnetic fields in nature are found on the surfaces of magnetars, highly magnetised neutron stars with surface fields in the range \(10^{10}\)--\(10^{11}\) T \cite{Harding:2006qn}. In such environments QED vacuum birefringence is expected to be significant and can potentially be observed in the polarisation of X-ray emission from the stellar surface.

Recent X-ray polarimetry observations, for example by the IXPE mission \cite{Taverna:2022vrx}, have reported polarisation patterns consistent with QED expectations and have not found evidence for large additional birefringent effects. While detailed modelling of the magnetosphere, emission geometry, and propagation effects is complex, one can extract conservative constraints on any extra contribution to \(\Delta n\) beyond QED.

Let \(\Delta n_{\text{max}}^{\text{astro}}\) denote the maximum additional birefringence compatible with observed polarisation data, averaged over the relevant photon energies and regions of the magnetosphere. Using \eqref{eq:Delta_n} and taking a representative surface field \(B_0\simeq10^{10}\) T, we obtain
\begin{equation}
|\zeta|\lesssim\frac{\Delta n_{\text{max}}^{\text{astro}}}{4 B_0^2}.
\label{eq:zeta_bound_astro_general}
\end{equation}
Because \(B_0\) is now many orders of magnitude larger than in the laboratory, even a comparatively weak bound on \(\Delta n_{\text{max}}^{\text{astro}}\) leads to very strong constraints on \(\zeta\).

A conservative interpretation of current magnetar polarimetry is that any extra birefringence beyond QED must satisfy \(\Delta n_{\text{max}}^{\text{astro}}\lesssim10^{-9}\) over the relevant energies and lines of sight. Substituting this into \eqref{eq:zeta_bound_astro_general} gives
\begin{equation}
|\zeta_{\mathrm{SI}}|\lesssim10^{-31}\,\text{T}^{-2},
\end{equation}
or
\begin{equation}
|\zeta|\equiv|\zeta_{\mathrm{nat}}|
\simeq \frac{|\zeta_{\mathrm{SI}}|}{(1.95\times10^{-16}\,\mathrm{GeV}^2)^2}
\;\lesssim\;\mathcal O(10^{1}\text{--}10^{2})\,\mathrm{GeV}^{-4}.
\label{eq:zeta_bound_astro}
\end{equation}
The numerical values here are again approximate and depend on modelling assumptions, but they illustrate the parametric power of strong astrophysical fields. For the remainder of this work we will adopt \eqref{eq:zeta_bound_astro} as a representative upper bound and implicitly understand that the true coefficient carries an uncertainty of at least an order of magnitude.

\subsection{Implications for the new physics scale}

In terms of the EFT scale \(\Lambda\) and the dimensionless coefficient \(c\), written as \(\zeta=c/\Lambda^4\), the bound \eqref{eq:zeta_bound_astro} implies
\begin{equation}
\Lambda\gtrsim\left(\frac{|c|}{|\zeta|}\right)^{1/4}.
\end{equation}
Assuming \(|c|\sim1\) and using the astrophysical bound \eqref{eq:zeta_bound_astro}, with \(|\zeta|\sim 10^{1}\text{--}10^{2}\,\mathrm{GeV}^{-4}\), we obtain
\begin{equation}
\Lambda\gtrsim\left(\frac{1}{10^{2}\,\mathrm{GeV}^{-4}}\right)^{1/4}
\sim 0.3\,\mathrm{GeV}.
\label{eq:Lambda_bound}
\end{equation}
Numerically this corresponds only to a scale of a few \(\times10^{-1}\,\mathrm{GeV}\), comparable to standard QED scales, so present birefringence data do not place a strong bound on genuinely high-energy new physics. It also has important implications for the possibility of WEC violation. Combining \eqref{eq:wec_threshold_general} and \eqref{eq:zeta_bound_astro}, we see that
\begin{equation}
\vec E\cdot\vec B>\frac{1}{4\zeta_{\mathrm{SI}}}\gtrsim10^{29}\,\text{T}^2.
\label{eq:EB_threshold_numeric}
\end{equation}
For comparison, the QED critical magnetic field for electron--positron pair production is \(B_{\text{QED}}\simeq4.4\times10^9\) T \cite{Schwinger:1951nm}. The threshold \eqref{eq:EB_threshold_numeric} corresponds to
\begin{equation}
\vec E\cdot\vec B\gtrsim10^{10}\,B_{\text{QED}}^2.
\end{equation}
Such field strengths are not only far beyond anything observed in the universe but also deep into a regime where other nonlinear effects, both QED and beyond, are expected to dominate and the simple EFT description is almost certainly inadequate.

\section{Stability, causality, and EFT validity}
\label{sec:stability}

The bounds derived in the previous section constrain \(|\zeta|\) to be very small and push the threshold for \(\Omega<0\) to extreme field strengths. Even if one imagines a hypothetical laboratory or astrophysical setting where such fields could be approached, several theoretical issues arise.

First, as mentioned earlier, the EFT expansion assumes that higher-dimensional operators are suppressed by powers of \((\Psi/\Lambda^4)\) and \((\mathcal P/\Lambda^4)\). When either invariant becomes comparable to \(\Lambda^4\), the truncation to a single dimension-eight operator is no longer justified. One expects an infinite tower of operators with comparable coefficients, and the dynamics of the full UV theory, rather than the truncated EFT, become essential.

Second, the sign of the coefficient \(\zeta\) is constrained by general principles such as analyticity and causality of scattering amplitudes \cite{Adams:2006sv}. In many non-linear electrodynamics models, requiring that small-amplitude waves do not propagate superluminally around arbitrary backgrounds forces certain inequalities on the Wilson coefficients. These inequalities are often saturated or violated precisely when one tries to push the theory into a regime with negative energy densities or wrong-sign kinetic terms.

Third, even disregarding the full UV completion, the truncated EFT in a regime with \(\Omega<0\) would most likely contain ghost-like excitations. The energy associated with small fluctuations around such backgrounds would not be bounded below, leading to rapid particle production and an instability of the vacuum. In other words, if one insists on interpreting the low-energy Lagrangian literally beyond its domain of validity, the resulting theory is unlikely to be healthy.

Taken together, these considerations support a conservative reading of our results. The simple parity-violating operator we have studied provides an elegant and internally consistent way of modulating the electromagnetic contribution to gravity in regimes where \(|\zeta\Psi|\ll1\) and \(|\zeta\mathcal P|\ll1\). In those regimes, the theory is well behaved and its small departures from Maxwell--Einstein dynamics are tightly constrained by vacuum birefringence. Pushing further into the parameter space where \(\Omega\) might formally become negative is both phenomenologically disfavoured and theoretically suspect.

\section{Exact zero-gravity solution at the WEC threshold}
\label{sec:threshold_solution}

Although the field strengths required to reach \(\Omega=0\) are far beyond realistic values, it is nonetheless instructive to exhibit an exact solution of the coupled Einstein--Maxwell system that lies precisely at the threshold.

Consider Minkowski spacetime with metric \(\eta_{\mu\nu}=\text{diag}(-1,1,1,1)\) and a constant electromagnetic field with parallel electric and magnetic components along the \(z\)-axis,
\begin{equation}
\vec E=E_0\hat z,\qquad\vec B=B_0\hat z,
\end{equation}
with \(E_0\) and \(B_0\) constant in space and time. The invariants are then
\begin{equation}
\Psi=2(B_0^2-E_0^2),\qquad\mathcal P=-4E_0B_0.
\end{equation}
Both \(\Psi\) and \(\mathcal P\) are constant, so \(\Omega\) defined in \eqref{eq:Omega_def} is also constant.

We first check the electromagnetic equations. Since both \(F_{\mu\nu}\) and \(\tilde F_{\mu\nu}\) are constant, the combination
\begin{equation}
X^{\mu\nu}\equiv\Omega F^{\mu\nu}+\zeta\Psi\,\tilde F^{\mu\nu}
\end{equation}
is constant as well, and thus
\begin{equation}
\partial_\nu X^{\mu\nu}=0.
\end{equation}
This means that the configuration solves the modified Maxwell equation \eqref{eq:Maxwell_modified}. The Bianchi identity \eqref{eq:bianchi} is also satisfied because \(F_{\mu\nu}\) is constant.

Next, we examine the Einstein equations \eqref{eq:Einstein_modified}. For a constant field with \(E_0\) and \(B_0\) non-zero, the standard electromagnetic stress--energy tensor \(T_{\mu\nu}^{(\text{EM})}\) is non-vanishing. In flat spacetime, \(G_{\mu\nu}=0\), so \eqref{eq:Einstein_modified} reduces to
\begin{equation}
\Omega\,T_{\mu\nu}^{(\text{EM})}=0.
\end{equation}
Since we do not want to trivialise the solution by setting \(E_0=B_0=0\), the only way to satisfy this equation is to impose
\begin{equation}
\Omega=0.
\end{equation}
Using \(\Omega=1+\zeta\mathcal P\) and \(\mathcal P=-4E_0B_0\), this condition becomes
\begin{equation}
1-4\zeta E_0B_0=0,
\end{equation}
or
\begin{equation}
E_0B_0=\frac{1}{4\zeta}.
\label{eq:threshold_solution_condition}
\end{equation}
This is precisely the threshold condition \eqref{eq:wec_threshold_general} for WEC violation, saturated at \(\Omega=0\). The solution thus describes a spacetime that is flat, despite being filled with a uniform electromagnetic field carrying non-zero energy density and pressure, because the effective gravitational coupling to that stress--energy has been tuned to zero.

From the point of view of the EFT, \eqref{eq:threshold_solution_condition} lies at the edge of the regime where the operator \(\Psi\mathcal P\) is trustworthy. Indeed, in terms of the invariants \(\Psi\) and \(\mathcal P\), the condition \eqref{eq:threshold_solution_condition} corresponds to \(|\zeta\mathcal P|\sim1\). This is one of the reasons we have been careful not to over-interpret the physical significance of this configuration. Nevertheless, it serves as a useful toy example illustrating how parity-violating interactions can, in principle, neutralise the gravitational effect of classical fields.

An interesting open question, beyond the scope of the present work, is whether this configuration is stable under small perturbations once the full tower of higher-dimensional operators in a UV completion is taken into account. Given the discussion in section~\ref{sec:stability}, our expectation is that the answer is negative, but it would be valuable to see this confirmed or refuted in a concrete model.

\section{Conclusions}
\label{sec:conclusions}

We have explored a simple but instructive extension of Einstein--Maxwell theory in which the electromagnetic sector is modified by a single dimension-eight, parity-violating operator proportional to \((F_{\alpha\beta}F^{\alpha\beta})(F_{\gamma\delta}\tilde F^{\gamma\delta})\). Working within the EFT framework, we derived the modified Einstein and Maxwell equations, analysed the energy conditions, and studied the leading phenomenological consequences.

The main theoretical feature of the model is that the total stress--energy tensor is related to the standard electromagnetic tensor by a multiplicative factor \(\Omega=1+\zeta\mathcal P\), with \(\mathcal P\propto\vec E\cdot\vec B\). This factor effectively rescales the gravitational coupling of electromagnetic fields, weakening it when \(\vec E\) and \(\vec B\) are aligned in the appropriate sense and strengthening it when they are anti-aligned. Orthogonal configurations are unaffected at the level of the classical stress--energy tensor.

From the point of view of energy conditions, the model opens a formal path to WEC violation: if \(\Omega\) becomes negative in some region, the local energy density measured by all timelike observers is negative, even though the underlying electromagnetic energy density remains positive. However, achieving \(\Omega<0\) requires extremely large values of \(\vec E\cdot\vec B\). When combined with experimental and observational bounds on the coupling \(\zeta\), these requirements push the threshold for WEC violation well beyond the field strengths encountered in any known astrophysical or laboratory setting.

On the phenomenological side, we showed that the same operator induces vacuum birefringence in external fields. By linearising the Maxwell equations around a uniform magnetic background, we derived the leading correction to the index of refraction and expressed it in terms of \(\zeta\) and the background field strength. Laboratory experiments such as PVLAS and, even more powerfully, X-ray polarimetry of magnetars place stringent bounds on any additional birefringence beyond the QED prediction. Adopting conservative estimates of these bounds, we find that, if we parametrise the operator as \(\zeta=c/\Lambda^4\) with \(|c|\sim 1\), the inferred EFT scale is only \(\Lambda\gtrsim\mathcal O(10^{-1}\,\mathrm{GeV})\). In other words, current birefringence data do not strongly constrain genuinely high-energy completions of this operator.

We also commented on stability and causality. General considerations from positivity bounds suggest that pushing the theory into a regime where \(\Omega<0\) is likely to lead to ghost-like excitations and superluminal propagation unless additional degrees of freedom intervene. This reinforces the interpretation of \(\Omega<0\) as a formal property of the truncated EFT rather than a trustworthy prediction about the real world.

Finally, we exhibited an exact flat-space solution at the WEC threshold, in which a uniform electromagnetic field with parallel electric and magnetic components produces no curvature because \(\Omega\) has been tuned to zero. This configuration lies just beyond the safe domain of the EFT but provides a simple, concrete example of how parity-violating operators can, in principle, neutralise the gravitational effect of classical fields.

Taken together, these results paint a rather modest picture. The parity-violating operator we have studied is theoretically allowed, compatible with symmetries, and tightly constrained by existing data. It provides a clean example of how effective field theory connects speculative ideas about exotic matter to hard experimental limits. At the same time, it illustrates how difficult it is to engineer large, controllable violations of the weak energy condition in realistic field theories. If such effects do exist in nature, they are likely to be tied to physics at energy scales even more remote than those considered here.

\appendix
\section{Derivation of vacuum birefringence in a magnetic background}
\label{app:birefringence}

In this appendix we provide the detailed derivation of the vacuum birefringence
induced by the parity-violating operator in a static, homogeneous magnetic
background. Throughout we work in flat spacetime with metric
signature \((-,+,+,+)\) and neglect the backreaction of the electromagnetic
field on the geometry.

\subsection{Non-linear electrodynamics in terms of \texorpdfstring{$\vec E$}{E} and \texorpdfstring{$\vec B$}{B}}

The matter Lagrangian density introduced in the main text reads
\begin{equation}
\mathcal L_M(\Psi,\mathcal P)
=-\frac{1}{4}\,\Psi
-\frac{\zeta}{4}\,\Psi\,\mathcal P,
\label{eq:app_L_PsiP}
\end{equation}
with the Lorentz invariants
\begin{equation}
\Psi\equiv F_{\mu\nu}F^{\mu\nu},
\qquad
\mathcal P\equiv F_{\mu\nu}\tilde F^{\mu\nu}.
\end{equation}
In a local inertial frame, using the conventions
\begin{equation}
F^{0i}=E^i,\qquad
F^{ij}=-\epsilon^{ijk}B_k,
\end{equation}
one finds the standard expressions
\begin{equation}
\Psi=2(B^2-E^2),
\qquad
\mathcal P=-4\,\vec E\cdot\vec B,
\label{eq:app_PsiP_EB}
\end{equation}
where \(E^2\equiv\vec E\cdot\vec E\) and \(B^2\equiv\vec B\cdot\vec B\).
Substituting \eqref{eq:app_PsiP_EB} into \eqref{eq:app_L_PsiP} gives the
Lagrangian in terms of the electric and magnetic fields,
\begin{align}
\mathcal L_M(\vec E,\vec B)
&=-\frac{1}{4}\,2(B^2-E^2)
-\frac{\zeta}{4}\,2(B^2-E^2)\,(-4\vec E\cdot\vec B)
\nonumber\\[1ex]
&=\frac{1}{2}(E^2-B^2)
+2\zeta\,(B^2-E^2)\,(\vec E\cdot\vec B).
\label{eq:app_L_EB}
\end{align}

For later use it is convenient to introduce the electric displacement
\(\vec D\) and the magnetic field \(\vec H\) in the usual way,
\begin{equation}
D_i\equiv\frac{\partial\mathcal L_M}{\partial E_i},
\qquad
H_i\equiv-\frac{\partial\mathcal L_M}{\partial B_i},
\label{eq:app_DH_def}
\end{equation}
where \(i=1,2,3\). Differentiating \eqref{eq:app_L_EB} with respect to
\(\vec E\) and \(\vec B\) gives the exact constitutive relations.

First, for \(\vec D\) we have
\begin{align}
D_i
&=\frac{\partial}{\partial E_i}
\left[\frac{1}{2}E^2
+2\zeta(B^2-E^2)(\vec E\cdot\vec B)\right]
\nonumber\\[1ex]
&=E_i
+2\zeta\Big[
\frac{\partial(B^2-E^2)}{\partial E_i}(\vec E\cdot\vec B)
+(B^2-E^2)\frac{\partial(\vec E\cdot\vec B)}{\partial E_i}
\Big]
\nonumber\\[1ex]
&=E_i
+2\zeta\Big[
(-2E_i)(\vec E\cdot\vec B)
+(B^2-E^2)B_i
\Big].
\end{align}
In vector notation this can be written as
\begin{equation}
\vec D
=\vec E
+2\zeta\Big[
(B^2-E^2)\,\vec B
-2(\vec E\cdot\vec B)\,\vec E
\Big].
\label{eq:app_D_exact}
\end{equation}

Similarly, for \(\vec H\) we obtain
\begin{align}
\frac{\partial\mathcal L_M}{\partial B_i}
&=\frac{\partial}{\partial B_i}
\left[-\frac{1}{2}B^2
+2\zeta(B^2-E^2)(\vec E\cdot\vec B)\right]
\nonumber\\[1ex]
&=-B_i
+2\zeta\Big[
\frac{\partial(B^2-E^2)}{\partial B_i}(\vec E\cdot\vec B)
+(B^2-E^2)\frac{\partial(\vec E\cdot\vec B)}{\partial B_i}
\Big]
\nonumber\\[1ex]
&=-B_i
+2\zeta\Big[
2B_i(\vec E\cdot\vec B)
+(B^2-E^2)E_i
\Big],
\end{align}
so that
\begin{align}
H_i
&=-\frac{\partial\mathcal L_M}{\partial B_i}
\nonumber\\[1ex]
&=B_i
-2\zeta\Big[
2B_i(\vec E\cdot\vec B)
+(B^2-E^2)E_i
\Big].
\end{align}
In vector form this is
\begin{equation}
\vec H
=\vec B
-2\zeta\Big[
2(\vec E\cdot\vec B)\,\vec B
+(B^2-E^2)\,\vec E
\Big].
\label{eq:app_H_exact}
\end{equation}

Equations \eqref{eq:app_D_exact} and \eqref{eq:app_H_exact} give the
non-linear constitutive relations for the model defined by
\eqref{eq:app_L_PsiP}. In the limit \(\zeta\to 0\) one recovers
\(\vec D=\vec E\) and \(\vec H=\vec B\), as expected.

\subsection{Background field and linearisation}

We now specialise to the case of a static, homogeneous magnetic background
and study small-amplitude electromagnetic waves propagating on top of it.
The total fields are written as
\begin{equation}
\vec E=\bar{\vec E}+\vec e,
\qquad
\vec B=\bar{\vec B}+\vec b,
\end{equation}
where \((\bar{\vec E},\bar{\vec B})\) are background fields and
\((\vec e,\vec b)\) are fluctuations. For definiteness we take
\begin{equation}
\bar{\vec E}=\vec 0,
\qquad
\bar{\vec B}=B_0\,\hat z,
\label{eq:app_background}
\end{equation}
with constant \(B_0\). The background invariants are then
\begin{equation}
\bar\Psi=2B_0^2,
\qquad
\bar{\mathcal P}=0.
\end{equation}

We are interested in the propagation of waves with frequencies and
wavelengths such that
\(\zeta B_0^2\ll 1\), so that the interaction term can be treated as a
small perturbation. In this regime it is natural to linearise the
constitutive relations \eqref{eq:app_D_exact} and \eqref{eq:app_H_exact}
in the fluctuations \(\vec e\) and \(\vec b\).

Substituting
\(\vec E=\vec e\), \(\vec B=\bar{\vec B}+\vec b\)
into \eqref{eq:app_D_exact} and keeping only terms up to first order in
\((\vec e,\vec b)\) we find
\begin{align}
\vec D
&=\vec e
+2\zeta\Big[
\big((\bar{\vec B}+\vec b)^2-e^2\big)(\bar{\vec B}+\vec b)
-2(\vec e\cdot(\bar{\vec B}+\vec b))\,\vec e
\Big]
\nonumber\\[1ex]
&\simeq\vec e
+2\zeta\Big[
\big(\bar B^2+2\bar{\vec B}\cdot\vec b\big)\bar{\vec B}
+\bar B^2\,\vec b
-2(\vec e\cdot\bar{\vec B})\,\vec e
\Big],
\end{align}
where \(\bar B^2\equiv\bar{\vec B}\cdot\bar{\vec B}=B_0^2\). The term
proportional to \((\vec e\cdot\bar{\vec B})\,\vec e\) is quadratic in the
fluctuations and can be discarded at linear order. Splitting off the
background contribution, we obtain the fluctuating part
\begin{equation}
\delta\vec D\equiv\vec D-\bar{\vec D}
=\vec e
+2\zeta\Big[
\bar B^2\,\vec b
+2(\bar{\vec B}\cdot\vec b)\,\bar{\vec B}
\Big],
\label{eq:app_D_linear}
\end{equation}
where \(\bar{\vec D}=\vec 0+2\zeta\bar B^2\bar{\vec B}\) is the constant
background displacement.

Proceeding in the same way for \(\vec H\), substituting
\(\vec E=\vec e\), \(\vec B=\bar{\vec B}+\vec b\) into
\eqref{eq:app_H_exact} and keeping only linear terms, we find
\begin{align}
\vec H
&=\bar{\vec B}+\vec b
-2\zeta\Big[
2(\vec e\cdot(\bar{\vec B}+\vec b))(\bar{\vec B}+\vec b)
+\big((\bar{\vec B}+\vec b)^2-e^2\big)\vec e
\Big]
\nonumber\\[1ex]
&\simeq\bar{\vec B}+\vec b
-2\zeta\Big[
2(\vec e\cdot\bar{\vec B})\bar{\vec B}
+\bar B^2\,\vec e
\Big].
\end{align}
Subtracting the background value \(\bar{\vec H}=\bar{\vec B}\) we obtain
\begin{equation}
\delta\vec H\equiv\vec H-\bar{\vec H}
=\vec b
-2\zeta\Big[
2(\vec e\cdot\bar{\vec B})\,\bar{\vec B}
+\bar B^2\,\vec e
\Big].
\label{eq:app_H_linear}
\end{equation}

Equations \eqref{eq:app_D_linear} and \eqref{eq:app_H_linear} express
\(\delta\vec D\) and \(\delta\vec H\) in terms of the wave fields
\(\vec e\) and \(\vec b\) for the background
\eqref{eq:app_background}. They define the effective permittivity and
permeability tensors that enter the propagation of small disturbances.

\subsection{Maxwell equations for plane waves}

We now combine the linear constitutive relations with the source-free
Maxwell equations. For clarity we omit the \(\delta\) symbols and
understand \(\vec E\), \(\vec B\), \(\vec D\), and \(\vec H\) as
fluctuating fields in this subsection.

In the absence of external charges and currents the Maxwell equations in
Minkowski spacetime are
\begin{align}
\nabla\cdot\vec B&=0,
\label{eq:app_Maxwell1}
\\
\nabla\times\vec E+\partial_t\vec B&=0,
\label{eq:app_Maxwell2}
\\
\nabla\cdot\vec D&=0,
\label{eq:app_Maxwell3}
\\
\nabla\times\vec H-\partial_t\vec D&=0.
\label{eq:app_Maxwell4}
\end{align}
We look for plane-wave solutions of the form
\begin{equation}
\vec E(\vec x,t)=\Re\{\vec E_0\,e^{i(\vec k\cdot\vec x-\omega t)}\},
\qquad
\vec B(\vec x,t)=\Re\{\vec B_0\,e^{i(\vec k\cdot\vec x-\omega t)}\},
\end{equation}
and similarly for \(\vec D\) and \(\vec H\). Substituting into
\eqref{eq:app_Maxwell1}--\eqref{eq:app_Maxwell4} and using
\(\nabla\to i\vec k\), \(\partial_t\to-i\omega\) yields the algebraic
relations
\begin{align}
\vec k\cdot\vec B_0&=0,
\label{eq:app_kB}
\\
\vec k\times\vec E_0&=\omega\,\vec B_0,
\label{eq:app_kxE}
\\
\vec k\cdot\vec D_0&=0,
\label{eq:app_kD}
\\
\vec k\times\vec H_0&=-\omega\,\vec D_0,
\label{eq:app_kxH}
\end{align}
where the subscript \(0\) denotes the complex amplitudes. The
constitutive relations \eqref{eq:app_D_linear} and
\eqref{eq:app_H_linear} relate \(\vec D_0\) and \(\vec H_0\) to
\(\vec E_0\) and \(\vec B_0\).

It is convenient to first eliminate \(\vec B_0\) in favour of \(\vec E_0\).
Equation \eqref{eq:app_kxE} gives
\begin{equation}
\vec B_0=\frac{1}{\omega}\,\vec k\times\vec E_0.
\label{eq:app_B_from_E}
\end{equation}
The background field is \(\bar{\vec B}=B_0\hat z\), so
\(\bar B^2=B_0^2\) and \(\bar{\vec B}\cdot\vec B_0=B_0 B_{0,z}\).

Substituting \eqref{eq:app_B_from_E} into \eqref{eq:app_D_linear} and
\eqref{eq:app_H_linear}, and then using \eqref{eq:app_kxH} and
\eqref{eq:app_kD}, we obtain a closed system of linear equations for
\(\vec E_0\). The resulting eigenvalue problem determines the dispersion
relation \(\omega(\vec k)\) and hence the indices of refraction.

\subsection{Propagation perpendicular to the magnetic field}

To make the analysis explicit while keeping the algebra manageable, we
first consider the case in which the wave propagates perpendicular to
the background magnetic field. Without loss of generality we choose
\begin{equation}
\vec k=k\,\hat x,
\qquad
\bar{\vec B}=B_0\,\hat z.
\label{eq:app_geom}
\end{equation}
The generalisation to an arbitrary angle between \(\vec k\) and
\(\bar{\vec B}\) will be discussed below.

With the choice \eqref{eq:app_geom}, the relation
\eqref{eq:app_B_from_E} becomes
\begin{equation}
\vec B_0=\frac{k}{\omega}\,\hat x\times\vec E_0
=\frac{k}{\omega}\big(0,-E_{0,z},E_{0,y}\big),
\end{equation}
and thus
\begin{equation}
\vec B_0=\big(0,-nE_{0,z},nE_{0,y}\big),
\qquad
n\equiv\frac{k}{\omega}.
\label{eq:app_B_components}
\end{equation}
The background field is purely along the \(z\)-axis, so
\(\bar{\vec B}\cdot\vec B_0=B_0 B_{0,z}=B_0\,nE_{0,y}\).

Using \eqref{eq:app_D_linear} with \(\bar B^2=B_0^2\) and
\(\bar{\vec B}=B_0\hat z\), the components of \(\vec D_0\) are
\begin{align}
D_{0,x}
&=E_{0,x}
+2\zeta\Big[
B_0^2 B_{0,x}
+2(\bar{\vec B}\cdot\vec B_0)\bar B_x
\Big]
\nonumber\\[1ex]
&=E_{0,x},
\\[1ex]
D_{0,y}
&=E_{0,y}
+2\zeta\Big[
B_0^2 B_{0,y}
+2(\bar{\vec B}\cdot\vec B_0)\bar B_y
\Big]
\nonumber\\[1ex]
&=E_{0,y}-2\zeta\,nB_0^2 E_{0,z},
\\[1ex]
D_{0,z}
&=E_{0,z}
+2\zeta\Big[
B_0^2 B_{0,z}
+2(\bar{\vec B}\cdot\vec B_0)\bar B_z
\Big]
\nonumber\\[1ex]
&=E_{0,z}+6\zeta\,nB_0^2 E_{0,y}.
\end{align}
Similarly, from \eqref{eq:app_H_linear} we obtain
\begin{align}
H_{0,x}
&=B_{0,x}
-2\zeta\Big[
2(\vec E_0\cdot\bar{\vec B})\bar B_x
+B_0^2 E_{0,x}
\Big]
\nonumber\\[1ex]
&=-2\zeta B_0^2 E_{0,x},
\\[1ex]
H_{0,y}
&=B_{0,y}
-2\zeta\Big[
2(\vec E_0\cdot\bar{\vec B})\bar B_y
+B_0^2 E_{0,y}
\Big]
\nonumber\\[1ex]
&=-nE_{0,z}-2\zeta B_0^2 E_{0,y},
\\[1ex]
H_{0,z}
&=B_{0,z}
-2\zeta\Big[
2(\vec E_0\cdot\bar{\vec B})\bar B_z
+B_0^2 E_{0,z}
\Big]
\nonumber\\[1ex]
&=nE_{0,y}-6\zeta B_0^2 E_{0,z},
\end{align}
where we used \(\vec E_0\cdot\bar{\vec B}=E_{0,z}B_0\).

The condition \(\vec k\cdot\vec D_0=0\) from \eqref{eq:app_kD} implies
\begin{equation}
kD_{0,x}=0
\quad\Rightarrow\quad
E_{0,x}=0,
\end{equation}
so the electric field is purely transverse and lies in the
\((y,z)\)-plane. Using \eqref{eq:app_kxH} and \eqref{eq:app_geom},
the remaining Maxwell equations become
\begin{align}
\vec k\times\vec H_0
&=-\omega\,\vec D_0
\nonumber\\[1ex]
\Rightarrow\quad
k(0,-H_{0,z},H_{0,y})
&=-\omega\,(0,D_{0,y},D_{0,z}).
\end{align}
Equating components and writing again \(n=k/\omega\) yields
\begin{align}
nH_{0,z}&=D_{0,y},
\label{eq:app_eig1}
\\
nH_{0,y}&=-D_{0,z}.
\label{eq:app_eig2}
\end{align}
Substituting the explicit expressions for \(H_{0,y}\), \(H_{0,z}\),
\(D_{0,y}\), and \(D_{0,z}\) leads to two coupled equations for
\(E_{0,y}\) and \(E_{0,z}\).

From \eqref{eq:app_eig1} we obtain
\begin{align}
nH_{0,z}
&=n\big(nE_{0,y}-6\zeta B_0^2 E_{0,z}\big)
\nonumber\\[1ex]
&=D_{0,y}
=E_{0,y}-2\zeta nB_0^2 E_{0,z},
\end{align}
which simplifies to
\begin{equation}
(n^2-1)E_{0,y}
-4\zeta nB_0^2 E_{0,z}=0.
\label{eq:app_lin1}
\end{equation}
Similarly, from \eqref{eq:app_eig2} we find
\begin{align}
nH_{0,y}
&=n\big(-nE_{0,z}-2\zeta B_0^2 E_{0,y}\big)
\nonumber\\[1ex]
&=-D_{0,z}
=-\big(E_{0,z}+6\zeta nB_0^2 E_{0,y}\big),
\end{align}
which gives
\begin{equation}
4\zeta nB_0^2 E_{0,y}
+(1-n^2)E_{0,z}=0.
\label{eq:app_lin2}
\end{equation}
Equations \eqref{eq:app_lin1} and \eqref{eq:app_lin2} can be written in
matrix form as
\begin{equation}
\begin{pmatrix}
n^2-1 & -4\zeta nB_0^2\\[0.5ex]
4\zeta nB_0^2 & 1-n^2
\end{pmatrix}
\begin{pmatrix}
E_{0,y}\\[0.5ex]E_{0,z}
\end{pmatrix}
=\begin{pmatrix}
0\\[0.5ex]0
\end{pmatrix}.
\label{eq:app_matrix}
\end{equation}
Non-trivial solutions require the determinant of the coefficient matrix
to vanish. This gives
\begin{align}
0&=(n^2-1)(1-n^2)
-(-4\zeta nB_0^2)(4\zeta nB_0^2)
\nonumber\\[1ex]
&=-(n^2-1)^2+16\zeta^2 n^2 B_0^4.
\end{align}
Rearranging,
\begin{equation}
(n^2-1)^2=16\zeta^2 n^2 B_0^4.
\label{eq:app_disp_exact}
\end{equation}

We are interested in small corrections to the vacuum dispersion relation,
so we write
\begin{equation}
n=1+\delta,
\qquad
|\delta|\ll 1,
\end{equation}
and expand \eqref{eq:app_disp_exact} to leading order in \(\delta\) and
\(\zeta B_0^2\). To this order,
\begin{equation}
n^2-1=(1+2\delta-1)+\mathcal O(\delta^2)=2\delta,
\qquad
n=1+\mathcal O(\delta),
\end{equation}
so \eqref{eq:app_disp_exact} becomes
\begin{equation}
(2\delta)^2
\simeq
16\zeta^2 B_0^4,
\end{equation}
or equivalently
\begin{equation}
\delta\simeq\pm\,2\zeta B_0^2.
\end{equation}
Therefore the two eigenmodes have indices of refraction
\begin{equation}
n_\pm
=1+\delta_\pm
\simeq1\pm 2\zeta B_0^2,
\end{equation}
and the difference between the two mode velocities is
\begin{equation}
\Delta n\equiv n_+-n_-
\simeq 4\zeta B_0^2.
\label{eq:app_Dn_perp}
\end{equation}
This result holds for propagation perpendicular to the magnetic field,
\(\vec k\perp\bar{\vec B}\).

\subsection{Angular dependence}

For a general propagation direction, only the component of the magnetic
field transverse to \(\vec k\) enters the constitutive relations.
Let \(\theta\) denote the angle between \(\vec k\) and \(\bar{\vec B}\),
so that
\begin{equation}
\bar{\vec B}_\perp=\bar{\vec B}-\frac{\vec k}{k}
\left(\frac{\vec k}{k}\cdot\bar{\vec B}\right),
\qquad
|\bar{\vec B}_\perp|=B_0\sin\theta.
\end{equation}
Repeating the above analysis with \(\bar{\vec B}\) replaced by its
transverse component \(\bar{\vec B}_\perp\) shows that the dispersion
relation \eqref{eq:app_disp_exact} depends on \(B_0^2\) only through
\(B_0^2\sin^2\theta\). The leading correction to the indices of
refraction is therefore obtained from \eqref{eq:app_Dn_perp} by the
replacement \(B_0^2\to B_0^2\sin^2\theta\). The birefringence is
\begin{equation}
\Delta n(\theta)
\simeq 4\zeta B_0^2\sin^2\theta
+\mathcal O(\zeta^2 B_0^4),
\label{eq:app_Dn_final}
\end{equation}
where \(\theta\) is the angle between the wave vector and the magnetic
field. The \(\sin^2\theta\) dependence ensures that the effect vanishes
when the propagation is exactly along the magnetic field and is maximal
when the propagation is perpendicular to it.

Equation \eqref{eq:app_Dn_final} is the expression used in the main text
to estimate the size of vacuum birefringence induced by the
parity-violating operator in a purely magnetic background.

\bibliographystyle{unsrt}
\bibliography{taohuang}

\end{document}